\documentclass[11pt,english,prl, article, a4paper]{revtex4}

\usepackage[T1]{fontenc}
\usepackage[latin1]{inputenc}
\usepackage{graphicx}

\makeatletter

\makeatletter

\usepackage{times}

\usepackage{setspace}

\usepackage{latexsym}

\usepackage{color}

\usepackage{graphics}

{\DeclareGraphicsExtensions{,.gz,.ps,.ps.gz,.jpg}}
\usepackage[dvips]{epsfig}

\usepackage{upgreek}

\usepackage{hyperref}

\makeatother

\usepackage{babel}
\makeatother
\begin{document}

\title{Self-organized adaptation of a simple neural circuit\\
enables complex robot behaviour}

\maketitle
{\Large \bigskip{}
 \bigskip{}
 }{\Large \par}

\begin{center}

Silke Steingrube$^{1,2}$, Marc Timme$^{1,3,4}$, Florentin
Wörgötter$^{1,4}$ and Poramate Manoonpong$^{1,4}$

\bigskip{}
 $^{1}$ Bernstein Center for Computational Neuroscience,\\
37073 Göttingen, Germany\\

$^{2}$ Department of Solar Energy, Institute for Solid
State Physics,\\ ISFH / University of Hannover, 30167 Hannover, Germany\\

$^{3}$ Network Dynamics Group, Max Planck Institute for Dynamics
\& Self-Organization,\\
37073 Göttingen, Germany\\

$^{4}$Faculty of Physics,University of Göttingen,\\
37077 Göttingen, Germany\\

Emails: silke@bccn-goettingen.de, timme@chaos.gwdg.de,
worgott@bccn-goettingen.de, poramate@bccn-goettingen.de

\bigskip{}

Ref: NPHYS-2009-04-00611B\\
File name: SelfOrganizedAdapationMainText.tex and
SelfOrganizedAdapationMainText.pdf

\end{center}

\bigskip{}
\bigskip{}

\newpage{}

\setcounter{page}{1}

\textbf{Controlling sensori-motor systems in higher animals or
complex robots is a challenging combinatorial problem, because
many sensory signals need to be simultaneously coordinated into a
broad behavioural spectrum. To rapidly interact with the
environment, this control needs to be fast and adaptive. Current
robotic solutions operate with limited autonomy and are mostly
restricted to few behavioural patterns. Here we introduce chaos
control as a new strategy to generate complex behaviour of an
autonomous robot. In the presented system, 18 sensors drive 18
motors via a simple neural control circuit, thereby generating
11 basic behavioural patterns (e.g., orienting, taxis,
self-protection, various gaits) and their combinations. The
control signal quickly and reversibly adapts to new situations
and additionally enables learning and synaptic long-term storage
of behaviourally useful motor responses. Thus, such neural control
provides a powerful yet simple way to self-organize versatile
behaviours in autonomous agents with many degrees of
freedom.\vspace{0.5cm}}

Specific sensori-motor control and reliable movement generation
constitute key prerequisites for goal-directed locomotion and
related behaviours in animals as well as in robotic systems. Such
systems need to combine information from a multitude of sensor
modalities and provide -- in real-time -- coordinated outputs to
many motor units \citep{Bernstein1967}. Already in relatively
simple animals, such as a common stick insect or a cockroach,
about 10 to 20 different basic behavioural patterns (several
different gaits, climbing, turning, grooming, orienting, obstacle
avoidance, attraction, flight, resting, etc.) arise from about ten
sensor modalities (e.g., touch sensors, vision, audition, smell,
temperature and vibration sensors) controlling on the order of 100
muscles. Nature apparently has succeeded in creating circuitries
specific for such purposes
\cite{Grillner2006,Bueschges2005,Pearson1984,Ijspeert2008} and
evolution has made it possible to solve the complex combinatorial
mapping problem of coordinating a large number of inputs and
outputs.

Conventional sensor-motor control methods for technical
applications do not yet achieve this proficiency. They typically
use for each behavioural output (e.g. each walking gait) one
specific circuit (control unit), the dynamics of which is
determined by several inputs. For example, one may decompose one
complex behaviour into a set of simple behaviours each controlled by
one unit (\cite{Brooks1986} "subsumption architecture"). In this
approach of behaviour-based robotics, sensors couple to actuators
in parallel. However, conventional methods are difficult to use in
self-organizing, widely distributed multi-input multi-output
systems \cite{Kurazume2002,Shkolnik2007}. For many such systems,
neural control appears more appropriate due to its intrinsically
distributed architecture and its capability to integrate new
behaviours \cite{Ijspeert2007,
Ishiguro2003,Buchli2006,Kuniyoshi2006,Arena2004,Kimura2007,Ayers,
Collins1994}.

Here, we address a complex high-dimensional coordination
problem employing one small neural circuit as a central pattern
generator (CPG). The goal is to generate different gaits in an
adaptive way and at the same time to coordinate walking with other
types of behaviours (such as orienting). To achieve this, the CPG circuit has an intrinsically chaotic dynamics similar
to that observed in certain biological central pattern generators
\cite{Rabinovich}. By means of a newly developed control method we
solve the conjoint problem of simultaneously detecting and
stabilizing unstable periodic orbits. The method is capable of
controlling many different periodic orbits in the same CPG, each
of which then leads to one specific activity pattern of the agent.
This happens in an autonomous and adaptive way because the states
of the sensory inputs of the agent at each moment determine which
period to control. As a consequence, the circuit can quickly adapt
to different situations. Followed by generic neural
postprocessing, this generates a wide range of specific behaviours
necessary to appropriately respond to a changing environment.
Furthermore, chaotic, uncontrolled dynamics proves behaviourally
useful, e.g., for self-untrapping from a hole in the ground.

\begin{figure}[!h]
\renewcommand{\figurename}{Fig.}
\epsfig{angle=0,width=15 cm,file= 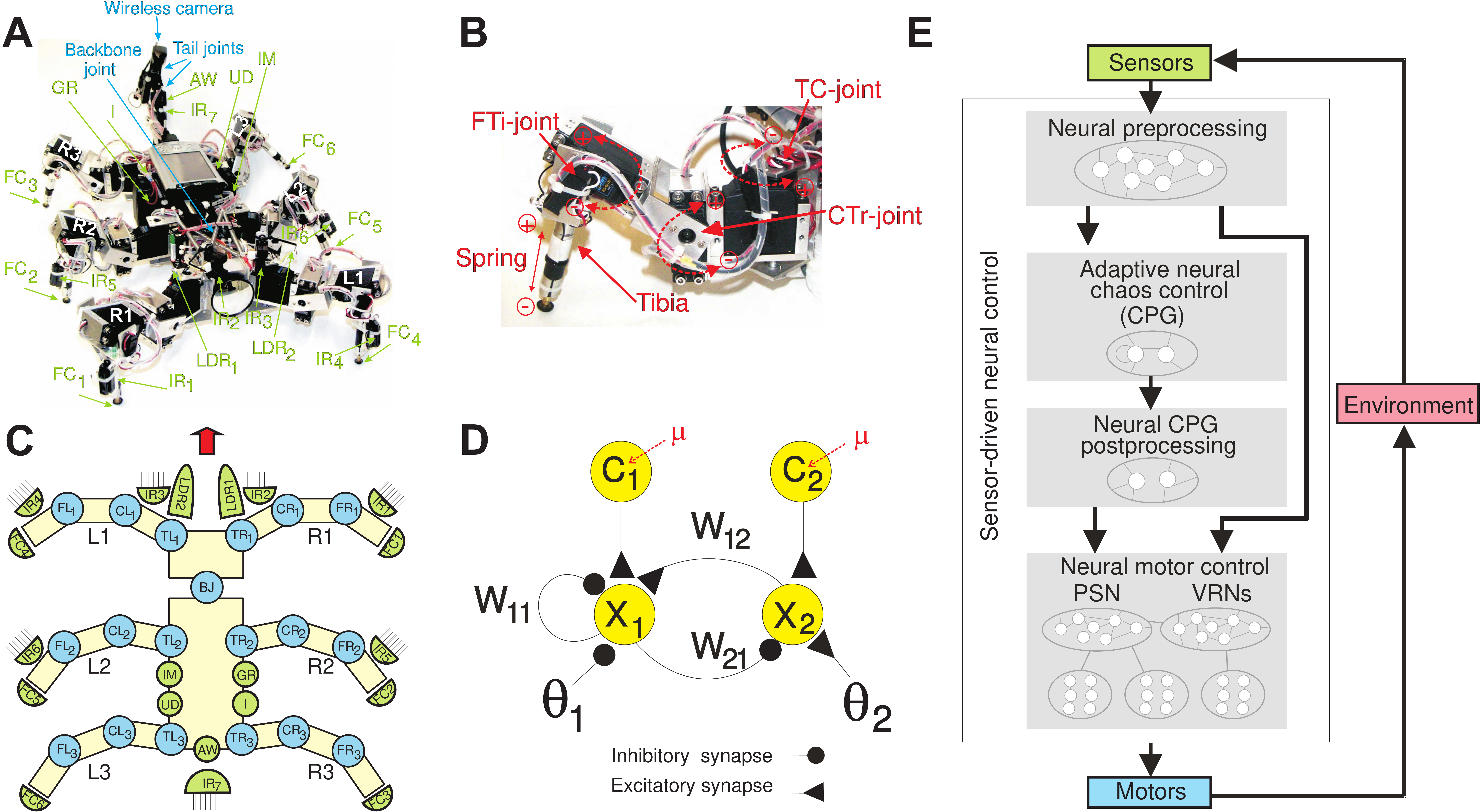}
\caption{{\footnotesize The six-legged walking machine AMOS-WD06
and the sensor-driven neural control setup.} \textbf{\footnotesize
(A) }{\footnotesize   AMOS-WD06 with 20 sensors (green arrows, 18
used here, IR sensors (IR$_{5,6}$) at the middle legs switched off
and not used (but see \cite{Manoonpong2008} for their
functionality)).} \textbf{\footnotesize (B) }{\footnotesize
Examples of joints at the right hind leg $R3$. Red-dashed arrows
show directions of forward (+)/backward ($-$) and up(+)/down($-$)
movements (see supplementary information and \textbf{Supplementary
Figure 1} for more details). TC-joint refers to the thoraco-coxal
joint for forward (+) and backward ($-$) movements. It corresponds
to TR$_{1,2,3}$ and TL$_{1,2,3}$ in \textbf{(C)}. The CTr-joint
refers to the coxa-trochanteral joint for elevation (+) and
depression ($-$) of the leg. The hexapod possesses six such
joints, three (CR$_{1,2,3}$) on its right and three (CL$_{1,2,3}$)
on its left, cf.~panel \textbf{(C)}. The FTi-joint refers to the
femur-tibia joint for extension (+) and flexion ($-$) of the
tibia. This corresponds to FR$_{1,2,3}$ and FL$_{1,2,3}$ in panel
\textbf{(C)}.} \textbf{\footnotesize (C) }{\footnotesize Scheme of
the hexapod AMOS-WD06 with 20 sensors (green), all 18 leg
motor-controlled joints and one backbone joint (blue).}
\textbf{\footnotesize (D) }{\footnotesize   Wiring diagram of the
neural control circuit (central pattern generator, CPG) consisting
of only two neurons with states $x_{i}$, $i\in \{ 1,2 \} $ (see eq. (\ref{eq1})) and
three recurrent synapses of strengths $w_{11}$, $w_{12}$, and
$w_{21}$. The $c_{i}$ are self-adapting control signals and $\mu$
is the control strength (see eqs. (\ref{eq2}), (\ref{eq3}),
(\ref{eq4}) and text for details). \textbf{\footnotesize (E)
}{\footnotesize   The setup of sensor-driven neural control for
stimulus induced behaviour of AMOS-WD06 (see text for functional
description and supplementary information and
\textbf{Supplementary Figure 2} for more details).}\label{amos}}}
\end{figure}

In addition to fast, reactive adaptation based on neural chaos
control (required to deal with sudden changes at sensor inputs),
the CPG-circuit introduced here allows also for learning on longer
time scales by synaptic plasticity. This way the system may also
permanently accommodate re-occurring correlations between sensor
inputs and motor outputs enabling the agent to gradually learn to
improve its behaviour.

As a prototypical example we consider a multi-sensor multi-motor
control problem of an artificial hexapod to create typical walking
patterns emerging in insects \cite{Wilson1966} as well as several
other behaviours. We solve two linked control problems for the
artificial hexapod AMOS-WD06 (\textbf{Fig.~\ref{amos}A, B})
\cite{Manoonpong2008}: sensor-driven gait selection
\cite{Orlovsky1999} and sensor-driven orienting behaviour
\cite{Manoonpong2008,Orlovsky1999}. For sensor-driven gait
selection, the system receives simultaneous inputs from thirteen
sensors (cf. \textbf{Fig.~\ref{amos}A, C}): two light-dependent
resistor sensors (LDR$_{1,2}$), six foot contact sensors
(FC$_{1,...,6}$), one gyro sensor (GR), one inclinometer sensor
(IM), one current sensor (I), one rear infra-red sensor (IR$_{7}$)
and one auditory-wind detector sensor (AW). They coact to
determine the dynamics of a very small, intrinsically chaotic
two-neuron module (described below) that serves as a central
pattern generator (CPG). After postprocessing, the CPG output
(\textbf{Fig.~\ref{amos}D, E}) selectively coordinates the action
of 18 motors into a multitude of distinct behavioural patterns.
Sensor-driven orienting behaviour is controlled via four additional
infra-red sensors (IR$_{1,2,3,4}$) together with the two
light-dependent resistor sensors (LDR$_{1,2}$) that generate
different types of tropism, e.g., obstacle avoidance (negative
tropism) and phototaxis (positive tropism) through two additional
standard (non-adaptive) neural subnetworks: one phase switching
network (PSN) and two identical modules of a velocity regulating
network (VRNs) (see \cite{Manoonpong2008} and supplementary
information for more details). In addition, one upside-down
detector sensor (UD) serves to activate a self-protective reflex
behaviour when the machine is turned into an upside-down position.
In the following, we describe the sensor-driven gait control
technique that is based on chaos control. The supplementary
information describes the technique of controlling sensor-driven
orienting behaviour.

\begin{figure}[!h]
\renewcommand{\figurename}{Fig.}
\epsfig{angle=0,width=15 cm,file= 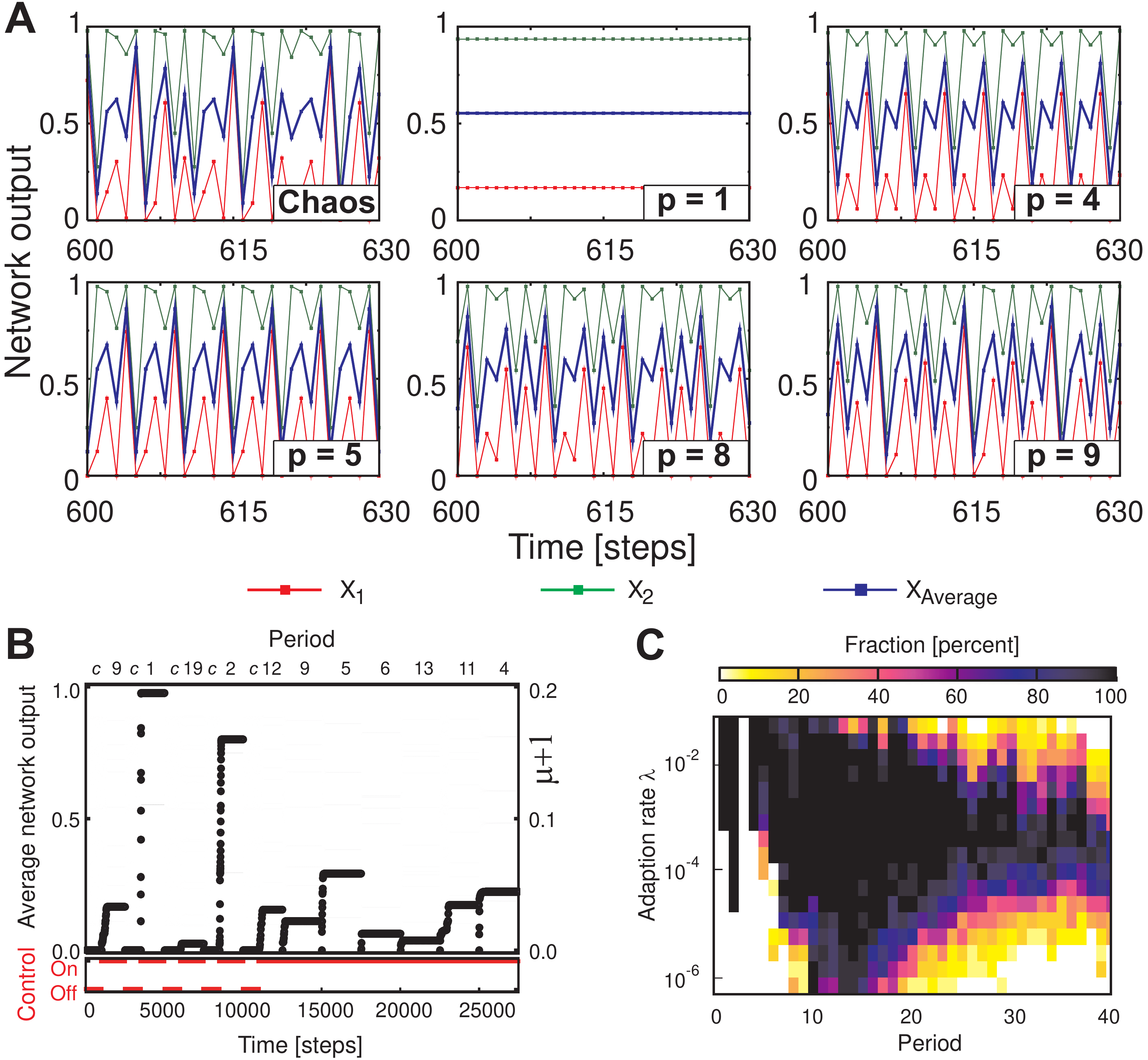}
\caption{{\footnotesize Control of unstable periodic orbits in the
chaotic CPG module.} \textbf{\footnotesize (A) }{\footnotesize CPG
dynamics without control (chaotic) and with control to specific
periodic orbits $p\in\{1,4,5,8,9\}$. Activity $x_{i}(t)$ of
neurons $i=1$ (red) and $i=2$ (green) are shown for some time
window $t\in[600,630]$ along with the average activity
$x_{av}$=$(x_{1}(t)+x_{2}(t))/2$ (blue).} \textbf{\footnotesize
(B) }{\footnotesize  Switching between different periodic orbits
(period indicated) and chaos ($c$) (adaption rate $\lambda$ =
0.05). The upper graph shows the average network output $x_{av}$
(thin dots, left axis) and control strength $\mu$ (thick dots, right
axis) for different target periods $p$. The lower graph shows the
time intervals of the control state (on/off). The target period is
changed every 2500 time steps (according to the top legend of
panel} \textbf{\footnotesize (B)}{\footnotesize ), while at the
same time the control strength $\mu$ is reset to $-1$. For the
first five target periods, control is intermediately switched off
for some time intervals such that the system exhibits chaotic
dynamics. For the final seven periods, control remains active such
that direct switching between periodic orbits occur with chaotic
dynamics only transiently. With increasing target periods, the
control strength tends to adapt to decreasing values $\mu$.}
\textbf{\footnotesize (C) }{\footnotesize  Fraction of correctly
controlled periods as a function of adaptation rate and period,
color coded from black (100\% correct) to white (0\% correct).
Every period is investigated for adaption rates in the range
$-$log $\lambda\in$\{ 1.2, 1.5,..., 6.3\} for 121 different random
initial conditions. An unstable periodic orbit of period three
apparently does not exist in the uncontrolled
dynamics.\label{PeriodicOrbits}}}
\end{figure}

To solve the combinatorially hard mapping problem of generating a
variety of gait patterns from multiple simultaneous inputs, we use
a simple module of two neurons $i\in\{1,2\}$
(\textbf{Fig.~\ref{amos}D}) as a CPG. The discrete time dynamics
of the activity (output) states $x_{i}(t)\in[0,1]$ of the circuit
satisfies

\begin{eqnarray}
x_{i}(t+1) & = &
\sigma\left(\theta_{i}+\sum_{j=1}^{2}w_{ij}x_{j}(t)+c_{i}^{(p)}(t)\right)\,\,\mbox{for}\,\,
i\in\{1,2\}\label{eq1}\end{eqnarray}

\noindent where $\sigma(x)=(1+\exp(-x))^{-1}$ is a sigmoid
activation function with biases $\theta_{i}$ and $w_{ij}$ is the
synaptic weight from neuron $j$ to $i$. The control signals
$c_{i}^{(p)}(t)$ act as additional biases that depend on a single
parameter $p$ only (the period of the output to be controlled) and
are uniquely determined by the sensory inputs. (cf. \textbf{Table
1}). We use synaptic weight and bias parameters (see
\textbf{Methods}) such that the circuit (eq. (\ref{eq1})) shows chaotic dynamics
if uncontrolled ($c_{i}^{(p)}(t)\equiv0$), see
\textbf{Fig.~\ref{PeriodicOrbits}A}.

\noindent In contrast to previous general methods of controlling
chaos \cite{Ott1993,Schmelcher1998} the method developed and
employed here both detects and stabilizes periodic orbits at the
same time and is implemented in a neural way. The signal
$c_{i}^{(p)}(t)$ is self-adapting and controls the dynamics of the
$x_{i}(t)$ to periodic orbits of period $p$ that are originally
unstable and embedded in the chaotic attractor, cf.
\cite{Pasemann2002, Ott1993,Heinz2005, Schoell2007,
Schimansky2007}. The fact that there is only one CPG makes the
control approach conceptually simple, easy to implement and, as
shown below, enables the system to self-adapt to new
combinations of sensory signals. Note, the combination of these
traits and their biological interpretation could not be so easily
achieved with any other pattern generation method (such as, for example, a
random-number generator). For a given period $p$, the control signal

\begin{equation}
c_{i}^{(p)}(t)=\mu^{(p)}(t)\sum_{j=1}^{2}w_{ij}\Delta_{j}(t)\label{eq2}\end{equation}

\noindent depends on the differences

\begin{equation}
\Delta_{j}(t)=x_{j}(t)-x_{j}(t-p)\label{eq3}\end{equation}

of states separated by one period $p$ and is applied every
$p+1$ time steps ($\Delta_j(t)=0$ and thus $c_{i}^{(p)}(t)=0$ at all other times) such that
each point of a periodic orbit is controlled sequentially. The control
strength $\mu^{(p)}$ adapts according to

\begin{equation}
\mu^{(p)}(t+1)=\mu^{(p)}(t)+\lambda\frac{\Delta_1^2(t)+\Delta_2^2(t)}{p}\label{eq4}\end{equation}

\noindent with adaption rate $\lambda$. The control strength is
initialized to $\mu(t_{\mbox{initial}})=-1$  whenever $p$ changes.
Here the scaling of the learning increment is heuristically chosen
as $1/p$ because a useful learning rate is found to decrease with
increasing period $p$.

\textbf{Figure~\ref{PeriodicOrbits}A} illustrates that the method
successfully generates distinct periodic orbits of different
periods, which in turn serve as CPG output patterns. Without
control, the CPG signal is chaotic. When being controlled, the CPG
dynamics reliably switches to one out of a large variety of
periodic outputs (\textbf{Fig.~\ref{PeriodicOrbits}B}) and control
is successful over a wide range of adaption rates
(\textbf{Fig.~\ref{PeriodicOrbits}C}). As the chaotic attractors
in various dynamical systems contain a large (often infinite)
number of unstable periodic orbits \cite{Ott1993, Heinz2005,
Schoell2007,Schimansky2007} it is in general possible to stabilize
many different periodic orbits in essentially any given
chaotically oscillating module that may then serve as a CPG. In
particular, the functionality is insensitive to variations in the
precise module dynamics and a specific type of CPG or a
multiple-unit CPG are not required.

\begin{figure}[htp]
\renewcommand{\figurename}{Fig.}
\epsfig{angle=0,width=15 cm,file= 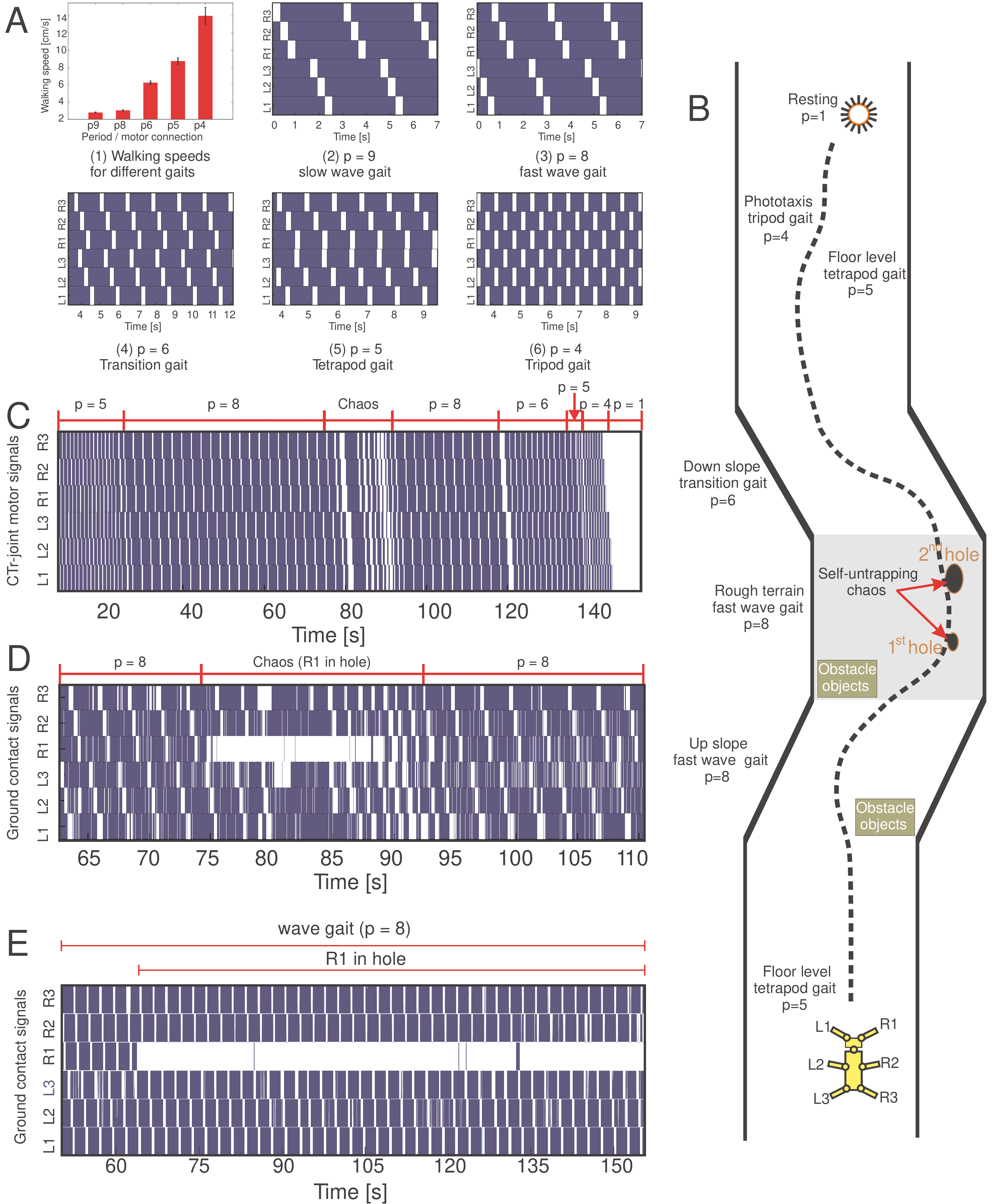}
\caption{{\footnotesize Chaos-controlled CPG generates
sensor-induced behavioural patterns of the hexapod AMOS-WD06.}
\textbf{\footnotesize (A) }{\footnotesize Examples of five
different gaits (see also \textbf{Supplementary Figure 4} and
\textbf{Video 1}) observed from the motor signals of the
CTr-joints (cf. \textbf{Fig. \ref{amos}B}) and walking speeds for
these gaits. Throughout the figure, blue areas indicate ground
contact or stance phase and white areas refer to no ground contact
during swing phase or stepping into a hole during stance phase.}
\textbf{\footnotesize (B) }{\footnotesize  Walking parcour of the
hexapod including barriers, obstacle objects, slopes, rough
terrain, holes in the ground and light source as phototropic
signal (\textbf{Supplementary Video 2}). Behavioural patterns and
associated periods of the CPG are indicated.}
\textbf{\footnotesize (C) }{\footnotesize  Gait patterns
(expressed as CTr-joint motor signals) observed during walking the
entire parcour (\textbf{Supplementary Video 2}).}
\textbf{\footnotesize (D) }{\footnotesize  Foot contact sensor
signals at time window 63 to 112}\emph{\footnotesize
s}{\footnotesize , indicating self-untrapping (foothold searching)
of right frontal leg ($R1$) as well as chaotic motion of other
legs. \textbf{\footnotesize (E) }{\footnotesize Without chaos,
untrapping is not successful, because a periodic gait does not
lift the leg out of the hole (compare \textbf{Supplementary Figure
6} and \textbf{Video 4}).} \label{HexapodBehaviour}}}
\end{figure}

Combining the adaptive neural chaos control circuit presented
above with standard PSN and VRNs postprocessing (cf. also
\textbf{Fig.~\ref{amos}E}) now enables sensor-driven control of a
large repertoire of behaviours. The extracted periodic orbits
generate the different gaits (\textbf{Fig.~\ref{HexapodBehaviour}}
and \textbf{Supplementary Video 1}), chaotic dynamics actively
supports un-trapping (cf. \textbf{Fig.~\ref{HexapodBehaviour}D} vs
\textbf{E}), and orienting behaviour arises simultaneously,
controlled by additional sensory inputs. These features enable the
robot to match environmental with behavioural complexity
(\textbf{Supplementary Video 2}); in particular, they create
specific targeted behaviours such as phototaxis (positive tropism)
and obstacle avoidance (negative tropism) (\textbf{Supplementary
Video 3}).

\textbf{Figure~\ref{HexapodBehaviour}A,B,C} exemplifies a sequence
of eight different behaviours (\textbf{Supplementary Video 2}):
standard walking in a tetrapod gait, up-slope walking in a wave
gait, rough-terrain walking in a wave gait, self-untrapping
through chaotic motion (\textbf{Supplementary Figure 6} and
\textbf{Video 4}), down-slope walking in a mixture gait (between
wave and tetrapod gait), active phototaxis by fast walking in a
tripod gait, and resting. As soon as obstacles are detected, the
machine moreover performs obstacle avoidance by turning
appropriately (\textbf{Supplementary Figure 5}). Here the
irregular chaotic 'ground state' of neural activity (cf.
\cite{Vreeswijk2006, Brunel2000, Zillmer2009, Jahnke2008,
Jahnke2009}) serves as an intermediate transient state that allows
for fast behavioural switching. As soon as the robot gets trapped
it actually operates chaotically and exploits chaos for efficient
untrapping (\textbf{Fig.~\ref{HexapodBehaviour}D}). This
demonstrates the capability of the robot to quickly alter its
behaviour in response to changing stimulus features from the
environment.

\begin{figure}[!h]
\renewcommand{\figurename}{Fig.}
\epsfig{angle=0,width=7.5 cm,file= 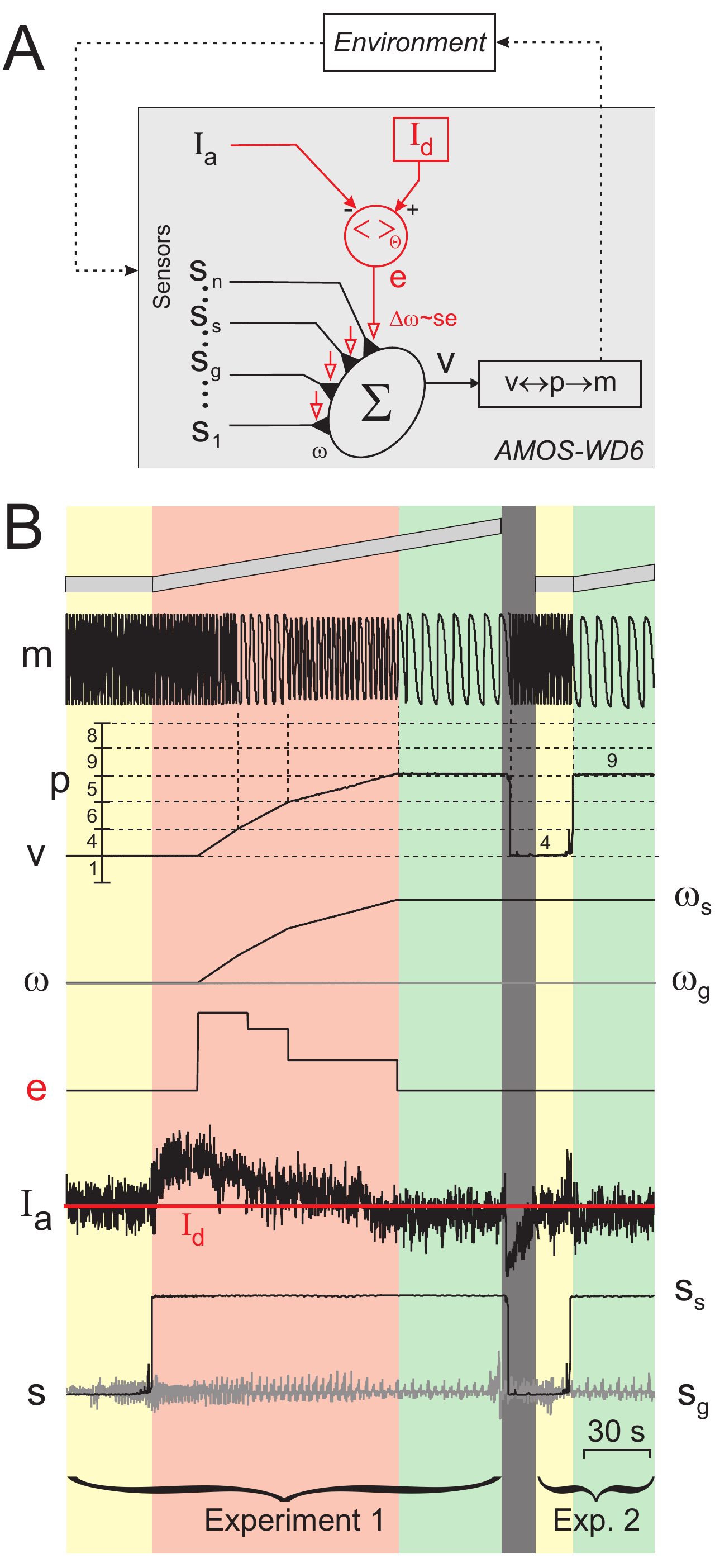}
\caption{{\footnotesize Learning sensor-motor mappings.
\textbf{(A)} Wiring diagram for learning. The learning circuit
is shown in red. A learning (summation) neuron $\sum$ produces
output $v$ from weighted sensor inputs $s_1\cdots s_n$. Black
triangles depict synapses. From output $v$ a gait $m$ is selected
using the CPG control signal $p$, leading to an average actual
motor current $I_a$ that depends on the terrain ("Environment"). The
actual motor current is compared with the stored default current
$I_d$ (red line, for tripod gait on flat terrain) creating an error signal
$e$, which is used for driving synaptic weight changes $\Delta
\omega$. The symbol $<>_\Theta$ denotes a thresholded averaging
process (see the Methods section). \textbf{(B)} Signals during two sequential
experiments (see also \textbf{Supplementary Video 5}). Colour code:
yellow = flat terrain, red = slope during learning, green = slope
after learning, grey = placement of robot back to starting
position. $m$ is the motor signal of a TC-joint;
$s_s$ is the inclinometer sensor signal; $s_g$ is the gyro sensor
signal. In the first experiment the robot on the slope learns to
choose the slow wave gait ($p=9$) that is energy saving and leads
to zero error and a drop of $I_a$. Only the correlated synapse
$\omega_s$ has grown; the other synapse $\omega_g$ remained close
to zero. In the second experiment triggering of the inclinometer
leads directly to the selection  of the slow wave gait without
further learning. Note $e$ is computed as an average, leading to a
delayed step function. The selection of $p$ from $v$ follows a
randomly chosen fixed mapping $v\leftrightarrow p$
shown by the dashed grid lines. Regardless of this
mapping, learning will always select the "zero-error gait" (here the slow wave
gait). \label{learnexp}}}
\end{figure}

The sensor-motor mapping so far was pre-assigned but can also
easily be learned (\textbf{Fig.~\ref{learnexp}A}). All artificial
CPGs built to date, including ours, directly map periodic gait
patterns ($p$) to motor patterns $m$. The most difficult open
problem here, thus, is to assure that periods $p$ are selected
appropriately given different sensory input conditions $s$, and
hence to learn a suitable mapping $s\rightarrow p$
(\textbf{Fig.~\ref{learnexp}A}). As the chaos-control strategy
uses only one single CPG, the learning problem becomes simple and
is solved using only one more single neuron that exhibits plastic
synapses. Plasticity is based on standard error minimization
learning, which we will describe in general terms next (for
details see \textbf{Methods}).

The state variable $v$ of the learning neuron linearly sums many
sensor inputs $s_k$ to $v=\sum_k \omega_k s_k$, where $\omega_k$
are the synaptic weights to be learned. We randomly assign periods
to neuron states in an arbitrary (but fixed) way $v\rightarrow p$
(\textbf{Fig.~\ref{learnexp}A}) such that different output levels
of $v$ result in different gaits. We will now discuss an example
where we use a steep and slippery slope on which the agent walks
upwards. Of all the agent's sensors, only the inclinometer $s_s$
(slope sensor) will be reliably triggered on the slope. Assuming
that its weight changes according to $d\omega_s /dt \sim s_s$, the
weight would grow gradually whenever a slope is sensed ($s_s>0$),
leading to increasing $v$ as long as the agent stays on the slope.
As the map $v\rightarrow p$ is fixed, the agent checks different
values of $p$ one by one trying out different gaits. As a
biologically motivated constraint, we now impose in addition that
the robot should choose to climb using an energy saving gait
\cite{Donald1981}. We hereby define a mechanism that stops
learning at that level of $v$, where such a gait is selected. This
is achieved by minimizing an error term $e$ that compares actual
energy uptake to the (low) energy uptake of the default gait on
flat terrain. If, while climbing, the agent chooses an energy
saving gait, this error will drop to zero. We, thus, modify our
learning rule to rely on the product of error and sensor signal,
$d\omega_s /dt \sim s_s\cdot e$, such that learning stops as soon
as the error is essentially zero. This happens when $\omega_s$
(and, thus, $v$) have grown to exactly the point where $p$ for the
lowest energy gait is selected.

\textbf{Figure~\ref{learnexp}B} illustrates the dynamics of this
learning experiment. Here the weight $\omega_s$ of the slope
sensor $s_s$ grows, whereas any uncorrelated synapse, e.g.,
$\omega_g$ from the gyro sensor $s_g$, remains unaffected
(\textbf{Fig.~\ref{learnexp}B}). This demonstrates that only the
relevant synapses learn. The output $v$ of the learning neuron
(\textbf{Fig.~\ref{learnexp}A}) follows these changes and
determines, via a threshold mechanism, different values of $p$
(\textbf{Fig.~\ref{learnexp}B}). As soon as $p$ selects the energy
saving slow wave gait (here $p = 9$), the error $e$ drops to zero,
stabilizing synapses and thereby fixing that gait. As the synaptic
values remain stored, the next time the hexapod encounters this
slope, the inclination sensor will immediately be triggered
leading to the same output $v$ and, hence, again to the selection
of the slow wave gait (\textbf{Fig.~\ref{learnexp}B}, right:
experiment 2).

In our single-CPG system learning is much simplified by the fact
that it only has to learn the single map $s\rightarrow p$. Thus,
the same neuron $v$ can also be used to learn other sensor-motor
mappings. For instance, in a second example of learning
(\textbf{Supplementary Figure 7 and Video 6}) we demonstrate how
the robot learns to escape from danger by choosing a particularly
fast gait.

Thus single-CPG control based on stabilizing unstable periodic
orbits enables self-adaptation of the required sensor-motor
mapping $s\leftrightarrow m$. This furthermore underlines a
central advantage of the single-CPG approach where pattern
generation is robust and learning becomes simple such that
additional sensor-motor conjunctions can also be implemented.

We have thus synthesized an integrated system, in which a small,
intrinsically chaotic CPG module brings together fast adaptivity
in response to changing sensor inputs with long term synaptic
plasticity. Both  mechanisms operate on the same network
components. The key ingredient here is the time-delayed feedback
chaos control that simultaneously detects and stabilizes the
dynamics of originally unstable periodic orbits in a biologically
inspired, neural way. It is capable of controlling a large number
of different periodic orbits of higher periods, a feature not
normally achieved in a robust way by standard time-delayed
feedback methods \cite{Schoell2007}. This finally permits
implementing learning in an efficient way, namely as a mode
selection process at the CPG.

As a consequence, the new strategy enables flexibly configurable
control that is readily implemented in hardware,
cf.~\cite{Manoonpong2008}. As it is based on controlling unstable
periodic orbits in a generic chaotic system, it does not
sensitively depend on the details of the dynamics. For instance,
the two-neuron architecture is not necessary and larger chaotic
circuits  work in a similar way. For the same reason, our strategy
may be generalized to integrate other behavioural patterns and can
also be applied for controlling different types of kinematic
(position controlled) walking machines and behaviours. Transfer to
dynamic walking \cite{Srinivasan2005} might be possible, too, but
would require adding control of additional state variables (e.g.
forces).

The chosen design is inspired by neural structures found in
insects. These combine adaptive CPG function \cite{Delcomyn1999}
with post-processing (\cite{Klaassen2002}, \cite{Asa2008}) similar
to the phase-switching network (PSN, \cite{Pearson1973}) and
velocity regulating network (VRN, \cite{Gabriel2007}) employed
here. Individual such network components had been used in earlier
studies and successfully provided partial solutions to artificial
motor control problems \cite{Ijspeert2007,
Ishiguro2003,Buchli2006,Kuniyoshi2006,Arena2004,Kimura2007,Ayers,
Collins1994} indicating that neural control is an efficient way
for solving complex sensori-motor control problems. For example,
Collins and Richmond \cite{Collins1994} have used a network of
four coupled nonlinear oscillators as hard-wired central pattern
generators to produce and switch between multiple quadrupedal gait
patterns by varying the network's driving signal and by altering
internal oscillator parameters. However, embodied control
techniques \cite{Pfeifer2007} for generating a variety of gait
patterns \cite{Srinivasan2005, Beer1997} jointly with other
sensor-driven behaviours \cite{Beer1997} in a system with many
degrees of freedom are still rare \cite{Ishiguro2003,
Kuniyoshi2006}.  Moreover, these systems either rely on only a
smaller number of sensors and motors, or, if more motors are
present \cite{Ijspeert2007}, their coordination forms
low-dimensional dynamics such as waves that constrain the motor
behaviour to snake- or salamander-like patterns with a uniform
gait. Both, small numbers of inputs and outputs and behavioural
restrictions reduce the sensor-motor coordination problem
substantially.

The capabilities of biological CPGs to generate chaotic as well as
periodic behaviour led to the hypothesis that chaos could serve as
a ground state for the generation of large behavioural repertoires
by the neural activity in these systems (for review see
\cite{Faure2003}). The current study now realizes this idea and
our chaos-based approach enables a complex combination of walking-
and orienting-behaviour. It simultaneously supports autonomous,
self-organized and re-configurable control by adaptively selecting
unstable periodic orbits from the chaotic CPG-module. Such CPGs
might moreover be used for  mutual entrainment between neural and
mechanical components of a behaving system \cite{Iida2006,
Pitti2006}.  Adding such features, however, would require further
investigations that are more system-specific.

Taken together this work suggests how a chaotic ground state of a
simple neuron module may be used in a versatile way for
controlling complex robots. It further demonstrates that chaos may
also play an active, constructive role for guiding the behaviour of
autonomous artificial as well as biological systems. The current
study still focuses on reactive motor behaviour. As periodic orbits
may be controlled also over longer periods of time, these systems
also offer the future possibility of implementing short term motor
memory. Decoupling the centralized control of the CPG from direct
sensor inputs would make it more persistent. This opens up the
opportunity of implementing behavioural components that make the
robotic system capable of navigating and moving with a certain
degree of memory-based planning and foresight \cite{Wehner2003,
McVea2007}.

\newpage{}

\textbf{Methods:}

\emph{Neural control}: Sensor-driven neural control for stimulus
induced walking behaviours consists of four neural modules: neural
preprocessing, adaptive neural chaos control (CPG), neural CPG
postprocessing, and neural motor control
(\textbf{Fig.~\ref{amos}E}). The controller acts as an artificial
perception-action system through a sensori-motor loop. All raw
sensory signals go to the neural preprocessing module. It consists
of several independent components which eliminate the sensory
noise and shape the sensory data (see supplementary information
for more details). The preprocessed light dependent resistor
(LDR$_{1,2}$), foot contact (FC$_{1,...,6}$), gyro (GR),
inclinometer (IM), and rear infra-red (IR$_{7}$) sensor signals
(\textbf{Fig.~\ref{amos}}) are transmitted to the adaptive neural
chaos control module. Simultaneously, other preprocessed infra-red
(IR$_{1,2,3,4}$), upsidedown detector (UD) as well as the
LDR$_{1,2}$ sensor signals (\textbf{Fig.~\ref{amos}}) are fed to
the neural motor control module.

In the adaptive neural chaos control module, a target period for
the chaos control is selected according to the incoming sensor
signals (see supplementary information). This module performs as a
CPG where its outputs for different periods determine the
resulting gait patterns of the machine (according to Table 1).
Here we set the bias values of the CPG circuit as
$\theta_{1}=-3.4$, $\theta_{2}=3.8$ and the three operating
synapses as $w_{11}=-22.0$, $w_{12}=5.9$, $w_{21}=-6.6$ ($w_{22} =
0.0$), such that it exhibits chaotic dynamics if uncontrolled
($c_{i}^{(p)}(t)\equiv0$), cf.
\textbf{Fig.~\ref{PeriodicOrbits}A}. The control strategy is
robust against changes of these parameters because it simply
relies on the CPG exhibiting chaotic dynamics. It is important to
note that chaos on the one hand serves as a ground state of the
CPG module, on the other hand it is also functionally used for
self-untrapping.

The CPG outputs are passed through the neural CPG postprocessing
module for shaping the signal that enters the neural motor control
module. The CPG postprocessing module is composed of two single
recurrent hysteresis neurons (more details in supplementary
information) which smooth the signals and two integrator units
which transform the discrete smoothed signals to continuous
ascending and descending motor signals. Finally, two fixed,
non-adaptive subnetworks, PSN and VRNs, of the neural motor
control module (\textbf{Supplementary Figure 6}) regulate and
change the CPG signals to expand walking capability allowing
turning as well as sidewards and backwards walking. In earlier
studies we have shown that the employed networks are robust within
a wide range of parameters \cite{Manoonpong2008}. In fact, it is
even possible to employ identical VRNs (without change in
structure or in parameters) in quadruped robots
\cite{Manoonpong2007} and transfer the PSN as well as the VRNs to
eight-legged machines \cite{Manoonpong2008}.

\emph{Learning}: Beyond sensor-driven neural control, we
additionally use a modified Widrow-Hoff rule \cite{Widrow1960} as
a learning mechanism to minimize energy consumption as a learning
goal (see supplementary information for other learning goals). We
define the output of the learning neuron as $v=\sum_k \omega_k
s_k$ and the rule as $d\omega_i/dt = \alpha \cdot e \cdot s_i$,
where $\alpha \ll 1$ is the learning rate. The error $e$ is given
as $e= <I_a-I_d>_{\Theta}$, the symbol $<~>$ denotes averaging
over 20 seconds and we set the error to zero if it is smaller than
$\Theta=0.01$. The variable $I_a$ is the currently used motor
current of all motors measured by a sensor
(\textbf{Fig.~\ref{amos}A,C}) and $I_d$ is the default current.
This is the average current used in a tripod gait on flat terrain.

\emph{Walking machine platform}: The six-legged walking machine
AMOS-WD06 is a biologically-inspired hardware platform. It
consists of six identical legs where each of them has three joints
(three degrees of freedom). All joints are driven by standard
servomotors. The walking machine has all in all 20 sensors
described in the main section where the potentiometer sensors of the
servomotors are not used for sensory feedback to the neural
controller. We use a Multi-Servo IO-Board (MBoard) to digitize
all sensory input signals and generate a pulse width modulated
(PWM) signal to control servomotor position. For the robot
walking experiments the MBoard is connected to a personal digital
assistant (PDA) on which the neural controller is implemented.
Electrical power supply is provided by batteries: one 7.4 V
Lithium polymer 2200 mAh for all servomotors, two 9 V NiMH 180 mAh
for the electronic board (MBoard) and the wireless camera, and
four 1.2 V NiMH 2200 mAh for all sensors (see supplementary for
more details).

\suppressfloats[ht!]
\begin{table}[h!]
\renewcommand{\tablename}{Table}
\caption{{\small List of different behaviours achieved given
environmental stimuli and conditions. "Default" means without
specific input signals. Note that the mapping between a gait and a
period is simply designed by using the fastest useful period,
which is p = 4 (p = 2 is too fast, p = 3 does not exist) for the
fastest gait and so on, where then p = 9 is the slowest gait.
Period p = 7 is in shape very similar to p = 6 and, therefore, it
is not used.}}
\begin{center}
\begin{tabular}{lll}
\hline\noalign{\smallskip} Environmental stimuli and conditions &
Period (p) & Behavioural pattern\\
\noalign{\smallskip} \hline \noalign{\smallskip}
Level floor                    & p = 5 &Tetrapod gait \\
Upward slope                   & p = 8 & Fast wave gait \\
Rough terrain (hole areas)                  & p = 8  & Fast wave gait \\
Losing ground contact                   & chaos  & Self-untrapping \\
Downward slope                     & p = 6  & Transition or mixture gait \\
Light stimuli                       & p = 4  & Tripod gait and orienting \\
& & toward stimuli\\
Strong light stimuli                       & p = 1  & Resting \\
Obstacles                        & p = 4,5,6,8, or, 9   & Orienting away from \\
& & stimuli\\
Turned upside-down                         & p = 4,5,6,8, or, 9  & Standing upside-down  \\
Attack of a predator                         & p = 4  & Tripod gait (escape behaviour)\\
Default        & p = 9  & Slow wave gait\\

\hline \label{table1}
\end{tabular}
\end{center}
\end{table}

\newpage{}

\bibliographystyle{apalike}

\vspace{1 cm}

\noindent \textbf{\large Acknowledgements}{\large \par}

\noindent We thank Frank Pasemann, Theo Geisel, Ansgar
B\"{u}schges, and Auke Jan Ijspeert for fruitful discussions and
acknowledge financial support by the Ministry for Education and
Science (BMBF), Germany, via the Bernstein Center for
Computational Neuroscience, grant numbers W3 {[}FW] and 01GQ0430
{[}MT] as well as by the Max Planck Society {[}MT]. FW
acknowledges funding by the European Commission
{}``PACO-PLUS\char`\"{}.

\noindent \textbf{\large Author contributions}{\large \par}

\noindent All authors conceived and designed the experiments,
contributed materials and analysis tools, and analyzed the data.
S.St. performed the numerical experiments. P.M. developed the
robotic system. P.M. and S.St. performed the robotic experiments.
M.T., F.W. and S.St. worked out the theory. M.T. and F.W.
supervised the numerical and robotic experiments. M.T., F.W. and
P.M. wrote the manuscript.

\noindent \textbf{\large Additional information}{\large \par}
\noindent Supplementary information accompanies this article on
www.nature.com/naturephysics. Reprints and permissions information
is available online at http://npg.nature.com/
reprintsandpermissions. Correspondence and requests for materials
should be addressed to P.M.

 \noindent \textbf{\large Competing
interests}{\large \par} \noindent The authors have declared that
no competing interests exist.

\end{document}